# Research on the Crystal Growth, Band Structure and Luminescence Mechanism of $(CH_3NH_3)_2HgI_4$

Linlin Liu[a,*], Zuanquan Chen[a], Wensi Yang[a], Sen Zhang[a], Jiaqi Zhu[a]

[a] *National Key Laboratory of Science and Technology on Advanced Composites in Special Environments, Harbin Institute of Technology, Harbin 150080, P.R. China*

[*]Corresponding author: 24S118261@stu.hit.edu.cn

## Abstract

Nuclear radiation detectors play a crucial role in fields such as nuclear safety and medical imaging. The core of their performance lies in the selection of detection materials. Semiconductor detectors have become a hot topic in current research due to their advantages such as small size, good energy resolution, and high detection efficiency. As one of the most promising materials for fabricating room - temperature nuclear radiation semiconductor detectors, $HgI_2$ exhibits excellent detection performance due to its high atomic number, large band gap, strong ray - stopping power, and high volume dark resistivity. However, issues such as poor chemical stability and low vacancy mobility of $HgI_2$ limit its development. Therefore, researchers have carried out inorganic doping/organic hybridization on it. By introducing the organic ligand $CH_3NH_3I$, the synthesis of organic - inorganic hybrid compounds based on $HgI_2$ is expected to significantly improve the stability of $HgI_2$. Research on organic - inorganic hybrid metal halide crystals shows that this material has great application potential in the field of luminescent materials.

In this paper, the crystal morphologies of $(CH_3NH_3)_2HgI_4$ under vacuum and in different solvent - anti - solvent systems were simulated. The crystal morphology of $(CH_3NH_3)_2HgI_4$ in the ethyl acetate - toluene system is long - plate - shaped, with the (002) plane being the largest exposed plane. Subsequently, a flake - shaped crystal with dimensions of 10 mm×9 mm×1 mm was prepared by the solution method in this system. XRD diffraction analysis revealed that its largest exposed plane is consistent with the morphology simulation results in this solvent system. Calculations show that $(CH_3NH_3)_2HgI_4$ is an indirect - band - gap semiconductor ($E_g$=3.28 eV). The ultraviolet - visible spectrum shows that its optical band gap is 3.11 eV, which is smaller than the calculated value. Therefore, it is speculated that there may be exciton luminescence in $(CH_3NH_3)_2HgI_4$. The calculations show that its first and second binding energy levels are 303.00 meV and 75.80 meV respectively. The existence of intrinsic peaks and exciton peaks was found in the room - temperature fluorescence spectrum, confirming the exciton luminescence phenomenon and indicating its wide application prospects in optoelectronic devices and the luminescence field.

# 1 Introduction

The discovery and development of nuclear radiation have not only promoted the progress of science and technology but also brought new challenges and threats. Nuclear radiation is a stream of microscopic particles released during the energy or structural transformation of atomic nuclei. This process involves various nuclear transitions and generates multiple radiation forms, including $\gamma$ rays, neutron radiation, and $\alpha$ particles [1]. These radiations not only reveal the mysteries of the atomic world but also pose potential threats to the environment and human health. To effectively detect harmful rays, detectors have become indispensable tools. They absorb the energy generated by radiation through specific materials and interact with the radiation to produce corresponding radiation signals. Subsequently, the signal collection equipment in the detector captures and processes these signals for subsequent analysis and research [2]. The research on detection materials is not only related to the basic working principle of detectors but also directly affects their performance and application scope.

Common detectors mainly include semiconductor detectors and gas detectors. Among them, semiconductor detectors are gradually becoming the focus of attention in the field of nuclear radiation detection due to their significant advantages, such as small volume, excellent energy resolution, wide application range, and high - efficiency detection ability. Silicon (Si) and germanium (Ge), as key semiconductor materials, are particularly suitable for low - energy radiation detection due to their narrow band gaps. In 1973, Malm et al. [3] studied the materials for room - temperature nuclear radiation semiconductor detectors and found that generally, the higher the atomic number, the better the detection efficiency, and $HgI_2$ and CdTe are more suitable materials for fabricating room - temperature semiconductor detectors compared to Si and Ge. In the subsequent extensive research and analysis of these two detection materials, $HgI_2$ showed relatively excellent performance. $HgI_2$ has many outstanding characteristics [4]: a high atomic number (Hg=70, I=53), a large crystal band gap (2.13 eV), high stopping power for X and $\gamma$ rays, and a high volume dark resistivity ($\rho>10^{13}$ $\Omega\cdot$cm), so it is recognized as one of the most promising new materials for preparing nuclear radiation detectors [5].

Under different environmental conditions, $HgI_2$ mainly exists in three phases: stable red $\alpha$-$HgI_2$ (tetragonal crystal system), metastable yellow $\beta$-$HgI_2^M$ (orthorhombic crystal system), and orange $HgI_2$. Among these phases, red $\alpha$-$HgI_2$ is the most thermodynamically stable state. However, the inherent limitations of $HgI_2$ crystals restrict its further development. Firstly, $HgI_2$ crystals are volatile in the atmosphere, especially the evaporation rate of iodine atoms is significantly higher than that of mercury atoms, resulting in easy degradation and deterioration of the crystal surface. This not only reduces the stability of the detector but also affects its detection performance [6]. In addition, from the perspective of luminescence characteristics, the luminescence of $HgI_2$ is highly temperature - dependent and can only exhibit luminescence at extremely low temperatures, which limits its application in a wider range. In view of these limitations and challenges of HgI$_2$, researchers have carried out extensive doping/hybridization studies to seek effective methods to optimize its performance, improve its stability, and expand its application scope, thus promoting the further development of HgI$_2$ materials. For the inorganic doping research of mercury - iodine - based crystals, some elements such as Br, Se, and S are added to the raw materials to form new compounds, improving their performance as detectors. However, in this process, heterogeneous atoms are inevitably introduced. Although these atoms enhance the hardness of the material, they also cause lattice distortion and increase brittleness. This poses a great challenge to subsequent crystal processing and limits its wide - scale promotion in practical applications. For the organic - inorganic hybridization research of mercury - iodine - based crystals, researchers use the unique structures and properties of organic ligands. Through the substitution of coordination elements, some iodine atoms in HgI$_2$ are successfully replaced by the coordination elements in the organic ligands. These new substances exhibit different properties and characteristics due to the differences in organic components. Therefore, the organic - inorganic hybridization of HgI$_2$ has become an effective means to synthesize new substances, explore new properties, and improve its performance as a detector, with great research value.

In this work, the organic ligand CH$_3$NH$_3$I was introduced to synthesize an organic - inorganic hybrid semiconductor material based on HgI$_2$. This approach improves the stability of HgI$_2$ through organic hybridization, overcomes its application limitations as a radiation - detection material, expands the application field of HgI$_2$, and enhances the stability of devices. During the hybridization process, the I provided by CH$_3$NH$_3$I coordinates and binds with Hg to form a stable [HgI$_4$]$^{2-}$ tetrahedral structure. In this process, CH$_3$NH$_3$I and HgI$_2$ are synthesized into (CH$_3$NH$_3$)$_2$HgI$_4$ in a molar ratio of 2:1, without destroying the basic structural unit of HgI$_2$. The organic matter plays a protective role around the [HgI$_4$]$^{2-}$ tetrahedron, enhancing the stability of the device. In addition, the steric - hindrance effect of the organic matter limits charge migration, forming a high exciton binding energy and providing conditions for stable excitons at room temperature. Therefore, the hybridization of CH$_3$NH$_3$I and HgI$_2$ generates a zero - dimensional structure of (CH$_3$NH$_3$)$_2$HgI$_4$. The reduction in its structural dimension may enable the material to exhibit stable exciton characteristics at room temperature, showing its potential as a new - type luminescent material.

## 2 Methods

In this work, the crystal morphology prediction and crystal growth of (CH$_3$NH$_3$)$_2$HgI$_4$ were carried out first. Crystal morphology prediction technology, as a key field in crystallization research, has made significant progress in recent years through molecular dynamics simulation software. Therefore, in this paper, a computational simulation analysis method was adopted to study the crystal morphology of (CH$_3$NH$_3$)$_2$HgI$_4$. In the simulation stage, the (CH$_3$NH$_3$)$_2$HgI$_4$ crystal was calculated using the Materials studio software. By applying the BFDH (Bravais -

Friedel - Donnay - Harker) method and the Growth Morphology method, the crystal planes that play a key role in the growth process of the $(CH_3NH_3)_2HgI_4$ crystal were found. In addition, the growth results of the $(CH_3NH_3)_2HgI_4$ crystal under different solvent conditions were simulated to explore the influence of solvents on the crystal morphology.

In this work, the $(CH_3NH_3)_2HgI_4$ crystal was prepared by the anti - solvent evaporation method in the solution method. During the crystal preparation process, considering that the crystal has good solubility in some organic solvents and is prone to decomposition reactions when encountering water, water - insoluble organic solvents should be selected. When the concentration of the solute in the solution exceeds its saturation point, crystals will gradually precipitate. In the process of preparing crystals by the solution method, the introduction of an anti - solvent can greatly reduce the solubility of the solute in the solution, so that it quickly transitions from an unsaturated state to a supersaturated state, which is more conducive to crystal growth. In addition, in order to more effectively regulate the concentration change of the solute in the solution and provide a relatively stable growth condition for it, a solvent with a slow evaporation rate should be selected as the anti - solvent. Considering the physical properties of the $(CH_3NH_3)_2HgI_4$ crystal, the anti - solvent should meet the following conditions:

(1) The anti - solvent should not react adversely with the target substance;

(2) The anti - solvent is immiscible with water;

(3) The anti - solvent and the main solvent must be miscible to ensure that the solvent can be effectively removed during the crystallization process;

(4) The evaporation rate of the anti - solvent is moderate, neither causing the crystal to grow too fast nor too slow;

(5) The anti - solvent should have a high solubility difference.

Based on the properties of some common organic solvents, as shown in Table 1, it can be found that $(CH_3NH_3)_2HgI_4$ is soluble in ethyl acetate, and the evaporation rate of ethyl acetate is medium among the alternative solvents. Therefore, ethyl acetate, methanol, etc. can be selected as solvents in this experiment. In the finally selected ethyl acetate system, ethyl acetate is soluble in benzene, toluene, chlorobenzene, etc., while $(CH_3NH_3)_2HgI_4$ is insoluble. Since the evaporation rate of benzene is similar to that of ethyl acetate, toluene and chlorobenzene can be selected as anti - solvents, that is, the ethyl acetate - toluene system and the ethyl acetate - chlorobenzene system.

Table 1 Properties of some organic solvents

| Solvent | Boiling Point/°C | Saturated Vapor Pressure （25°C）/kPa | Solubility of $(CH_3NH_3)_2HgI_4$ |
| --- | --- | --- | --- |
| Methanol | 64.7 | 16.83 | Soluble |
| Dimethylformamide | 153 | 0.533 | Soluble |
| Ethyl acetate | 77 | 12.617 | Soluble |
| Benzene | 80 | 12.691 | Insoluble |
| Toluene | 111 | 3.792 | Insoluble |
| Xylene | 138 | 1.168 | Insoluble |
| Chlorobenzene | 132 | 1.596 | Insoluble |

## 3 Results

Based on the given unit cell structure parameters [7], as shown in Table 2, a unit cell model of the $(CH_3NH_3)_2HgI_4$ crystal was constructed, as shown in Figure 1(a). This crystal belongs to the orthorhombic crystal system with the Pbca space group. Its lattice constants are $a$=10.985 Å, $b$=12.261 Å, $c$=21.203 Å, and the angles between the three crystal axes are all right angles ($\alpha=\beta=\gamma=90°$).

In the $(CH_3NH_3)_2HgI_4$ crystal, the $[HgI_4]^{2-}$ tetrahedron constitutes its basic structural unit. However, these tetrahedrons are surrounded by $CH_3NH^{3+}$ ions, which hinder the three-dimensional expansion of the $[HgI_4]^{2-}$ tetrahedron in the crystal, thus forming zero-dimensional structural characteristics.

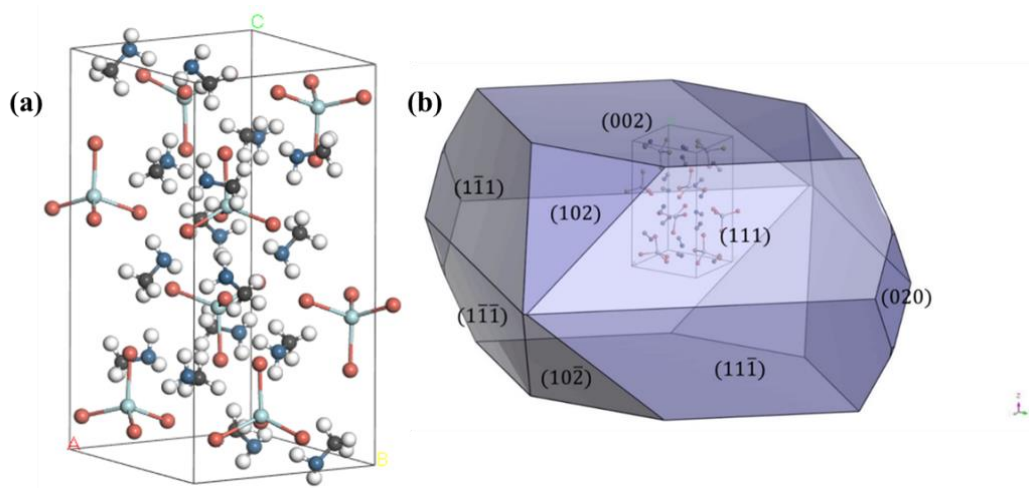

Figure 1 (a) Unit cell structure of $(CH_3NH_3)_2HgI_4$; (b) Morphology of $(CH_3NH_3)_2 HgI_4$ crystal simulated by the BFDH method

Table 2 Unit cell parameters of $(CH_3NH_3)_2HgI_4$

| Unit call | Space group | $a$/Å | $b$/Å | $c$/Å | $\alpha$/° | $\beta$/° | $\gamma$/° |
|---|---|---|---|---|---|---|---|
| $(CH_3NH_3)_2HgI_4$ | $P$bca | 10.985 | 12.261 | 21.203 | 90 | 90 | 90 |

Through the simulation by the BFDH method, the morphology of the $(CH_3NH_3)_2HgI_4$ crystal under vacuum is polyhedral granular, as shown in Figure 1(b). The parameters of its important crystal planes are shown in Table 3. It can be inferred that the importance order of the crystal planes is (002) > (111) > (102) > (020) > (021) > (200). Crystal planes with larger areas are more likely to be exposed during the growth process and thus have higher importance. Among all the simulated crystal planes, the (002) crystal plane is regarded as the main morphological crystal plane due to its large area, and it has the most significant impact on the overall morphology of the crystal. In contrast, the (020) and (021) crystal planes are likely to disappear during the growth process due to their small areas and are easily affected by other factors.

Table 3 Parameters of each crystal plane of the $(CH_3NH_3)_2HgI_4$ crystal under vacuum

| (hkl) | %Total facet area |
| --- | --- |
| （002） | 28.744 |
| （111） | 58.605 |
| （102） | 11.090 |
| （020） | 1.538 |
| （021） | 0.023 |

Based on the Materials Studio (MS) software, the geometry of the primitive cell was optimized using the Forcite module to ensure the stability of the structure. Subsequently, the crystal planes were cut according to the previously determined order of the main crystal planes. In order to simulate the adsorption behavior of organic solvent molecules on these crystal planes, corresponding models were established. To meet the requirements of molecular dynamics simulation for a large - data sampling space, these crystal planes were expanded into a 3×3×3 periodic supercell, and a vacuum layer with a thickness of 20 Å was set to simulate the interaction between the crystal plane and solvent molecules in the actual environment.

When predicting the crystal morphology, we usually rely on the crystal plane attachment energy of each crystal plane as an important indicator. The calculation formula for the crystal plane attachment energy $E_{att}$ is as follows [8]:

$$E_{att} = E_{latt} - E_{slice} \quad (1)$$

where $E_{att}$ represents the lattice energy, that is, the energy released during the crystal formation process; and $E_{slice}$ represents the two - dimensional lattice energy of the crystal plane. It should be noted that the positive and negative values of the crystal plane attachment energy $E_{att}$ can directly reflect whether the crystal plane absorbs or releases energy during the adsorption process.

Since the adsorption of the solvent on the crystal surface will affect its crystal plane

attachment energy, the above formula is modified. The modified solvent attachment energy is:

$$E'_{att} = E_{att} - \frac{A_{acc}}{A_{box}} \times [E_{tot} - E_{cry} - E_{sol}] \quad (2)$$

where $A_{acc}$ is the accessible area of the solvent, $A_{box}$ is the cross-sectional area of the solvent box, $E_{tot}$ is the total energy of the crystal-solvent bilayer structure, $E_{cry}$ is the energy of the crystal layer, and $E_{sol}$ is the energy of the solvent layer.

The energies of each crystal plane were calculated by the MS software, and the crystal plane adsorption energies in different solvent and anti-solvent systems were calculated by the above formula (2), as shown in Tables 4, 5, and 6.

Table 4 The crystal plane layer, solvent layer, and modified attachment energies of $(CH_3NH_3)_2HgI_4$ with ethyl acetate solvent molecules

| Crystal Plane (hkl) | $E_{att}$ Kcal/mol | $E_{total}$ Kcal/mol | $E_{solv}$ Kcal/mol | $E_{sury}$ Kcal/mol | $E_{int}$ Kcal/mol | $S\frac{A_{hkl}}{A_{box}}$ | $E'_{att}$ Kcal/mol |
|---|---|---|---|---|---|---|---|
| (001) | -28.55254445 | -323.386 | 3773.729 | -342.388 | -3754.727 | 0.385 | -26.566 |
| (111) | -40.56767086 | -524.083 | 974814541.969 | -579.923 | -974814486.1 | 0.326 | -19.185 |
| (102) | -40.24332064 | -395.042 | 542275282.486 | -517.713 | -542275159.8 | 0.384 | -29.759 |
| (020) | -57.11673428 | -294.900 | 1830681.746 | -210.965 | -1830765.681 | 0.382 | -13.423 |
| (021) | -58.41136575 | -321.362 | 12098937642.421 | -371.675 | -12098937592 | 0.374 | 1.315 |
| (200) | -59.68973474 | -163.437 | 6821391.107 | -184.038 | -6821370.506 | 0.338 | 48.578 |

Table 5 The crystal plane layer, solvent layer, and modified attachment energies of $(CH_3NH_3)_2HgI_4$ in the ethyl acetate - toluene solvent system

| Crystal Plane (hkl) | $E_{att}$ Kcal/mol | $E_{total}$ Kcal/mol | $E_{solv}$ Kcal/mol | $E_{sury}$ Kcal/mol | $E_{int}$ Kcal/mol | $S\frac{A_{hkl}}{A_{box}}$ | $E'_{att}$ Kcal/mol |
|---|---|---|---|---|---|---|---|
| (001) | -28.55254445 | -246.584 | 727.929 | -342.388 | -632.125 | 0.385 | 214.8155806 |
| (111) | -40.56767086 | -325.501 | 178057959.826 | -579.923 | -178057705.4 | 0.326 | 58046771.39 |
| (102) | -40.24332064 | -317.936 | 99034104.563 | -517.713 | -99033904.79 | 0.384 | 38028979.19 |
| (020) | -57.11673428 | -80.833 | 336711.041 | -210.965 | -336580.909 | 0.382 | 128516.7905 |
| (021) | -58.41136575 | -149.248 | 2210291844.598 | -371.675 | -2210291622 | 0.374 | 826649008.3 |
| (200) | -59.68973474 | -12.086 | 1247528.262 | -184.038 | -1247356.31 | 0.338 | 421546.743 |

Table 6 The crystal plane layer, solvent layer, and modified attachment energies of $(CH_3NH_3)_2HgI_4$ in the ethyl acetate - chlorobenzene solvent system

| Crystal Plane (hkl) | $E_{att}$ Kcal/mol | $E_{total}$ Kcal/mol | $E_{solv}$ Kcal/mol | $E_{sury}$ Kcal/mol | $E_{int}$ Kcal/mol | $S\frac{A_{hkl}}{A_{box}}$ | $E'_{att}$ Kcal/mol |
|---|---|---|---|---|---|---|---|
| (001) | -28.55254445 | -246.584 | 727.929 | -342.388 | -632.125 | 0.385 | 214.8155806 |
| (111) | -40.56767086 | -325.501 | 178057959.826 | -579.923 | -178057705.4 | 0.326 | 58046771.39 |
| (102) | -40.24332064 | -317.936 | 99034104.563 | -517.713 | -99033904.79 | 0.384 | 38028979.19 |
| (020) | -57.11673428 | -80.833 | 336711.041 | -210.965 | -336580.909 | 0.382 | 128516.7905 |
| (021) | -58.41136575 | -149.248 | 2210291844.598 | -371.675 | -2210291622 | 0.374 | 826649008.3 |
| (200) | -59.68973474 | -12.086 | 1247528.262 | -184.038 | -1247356.31 | 0.338 | 421546.743 |

The morphology diagrams of the $(CH_3NH_3)_2HgI_4$ crystal in different solvent systems were obtained, as shown in Figure 2. Figure 2(a) shows the crystal morphology in the ethyl acetate solvent - anti - solvent system. It can be seen from the crystal morphology diagram that the crystal obtained in the ethyl acetate system is flake - shaped, which is more suitable for the large - area crystal growth of $(CH_3NH_3)_2HgI_4$. In the ethyl acetate - toluene system, it is long - plate - shaped, and its largest exposed crystal plane is the (102) crystal plane. In the ethyl acetate - chlorobenzene system, it is long - flake - shaped, and the largest exposed plane is the (002) plane.

During the crystal growth process, the introduction of the anti - solvent significantly affects the crystal morphology. This influence not only stems from the raw material components, the interaction between the anti - solvent and the solvent, and the properties of the anti - solvent itself (such as boiling point, polarity, etc.), but is also strongly regulated by the crystallization experimental conditions (such as the amount, addition rate, and position of the anti - solvent). The anti - solvent promotes the formation of precipitation by reducing the solubility of the solute in the solution and directly affects the final morphology of the crystal. Although the influence mechanism of the anti - solvent on the crystal morphology has not been clearly explained, one of the reasons is the strong interaction between the solvent and the anti - solvent, especially the hydrogen - bond interaction caused by the inherent polarity of the solvent used, so the system is forced to crystallize. In addition, the functional groups of the anti-solvent molecules also play an important role in regulating the crystal morphology. Therefore, it is inferred that when toluene, an anti-solvent, is added to the ethyl acetate solvent system, the crystal size will become smaller due to their interaction. If an anti-solvent with a higher polarity is added, an even smaller crystal will be obtained.

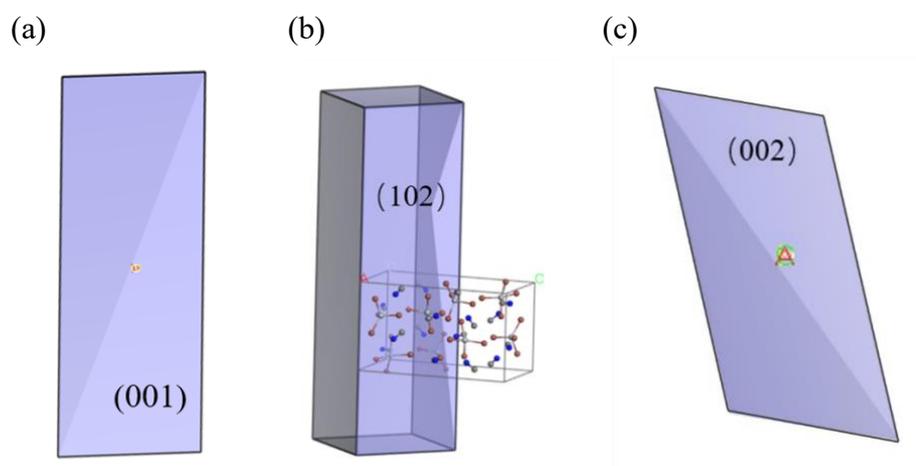

Morphology prediction diagrams of $(CH_3NH_3)_2HgI_4$ under different solvents: (a) Ethyl acetate (b) Ethyl acetate - toluene (c) Ethyl acetate – chlorobenzene

Table 5 The area percentages of important crystal planes in the morphology diagrams of $(CH_3NH_3)_2HgI_4$ crystals grown in different solvent systems

| Crystal Plane (hkl) | Ethyl Acetate %Total facet area | Ethyl acetate - toluene %Total facet area | Ethyl acetate – chlorobenzene %Total facet area |
|---|---|---|---|
| (001) | 99.73659982 | | 99.66630483 |
| (111) | | | |
| (102) | | 86.66668901 | |
| (020) | 0.20210128 | | 0.33369517 |
| (021) | | 13.33331099 | |
| (200) | 0.06129889 | | |

After the crystal morphology prediction, crystal growth was carried out by the solution method. The selected ethyl acetate solvent - anti-solvent system was used for the experiment. The experimental steps are shown in Figure 3, and the detailed steps are as follows:

(1) Weigh 1.001 g of MAI and 1.434 g of $HgI_2$ according to a molar ratio of 2:1. $HgI_2$ is a red powder, and MAI is white granular.

(2) Prepare a glass beaker, add 14 mL of ethyl acetate, and dissolve the weighed chemicals in ethyl acetate. Stir well with a magnetic stirrer at room temperature to ensure they are fully dissolved in the ethyl acetate solvent.

(3) Filter the above - mentioned mixed solution with a 0.45 - μm filter head to obtain a clear mixed solution. Add 12 mL of the anti - solvent drop - by - drop to the clear mixed solution, and stir well to obtain a transparent yellow solution.

(4) After filtering, seal the mouth of the small beaker with plastic wrap and make one or two small holes in the plastic wrap to facilitate the volatilization of the solvent. Place it in a quiet and dry environment for slow crystallization.

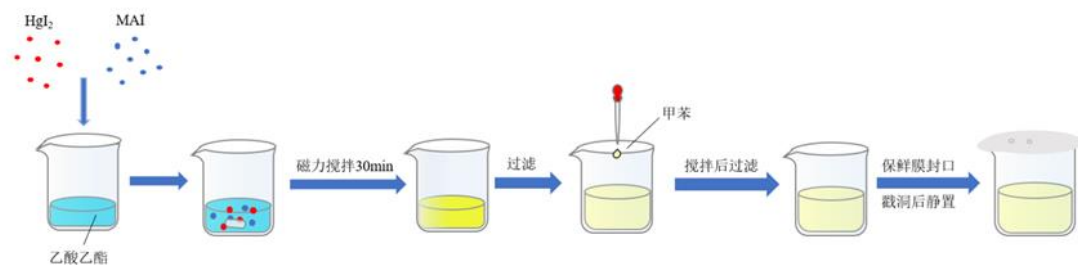

Figure 4 Experimental steps diagram

In the ethyl acetate - toluene system, the macroscopic growth result of the $(CH_3NH_3)_2HgI_4$ crystal is shown in Figure 4(a). It can be seen that the crystal obtained in the ethyl acetate - toluene system is a yellow - blocky crystal. The total raw material is 2.435 g, and the product with a size greater than 1 mm is 17.18 g. The yield is calculated to be 89.74%. One of the crystals with dimensions of approximately 10 mm×9 mm×1 mm has a regular morphology, and currently, the ethyl acetate - toluene system is the most ideal system in this experimental preparation. In the ethyl acetate - chlorobenzene system, the macroscopic growth result of the $(CH_3NH_3)_2HgI_4$ crystal is shown in Figure 4(b). The $(CH_3NH_3)_2HgI_4$ crystal is a yellow - transparent needle - shaped solid as a whole, with dimensions of approximately 10 mm×3 mm×1 mm.

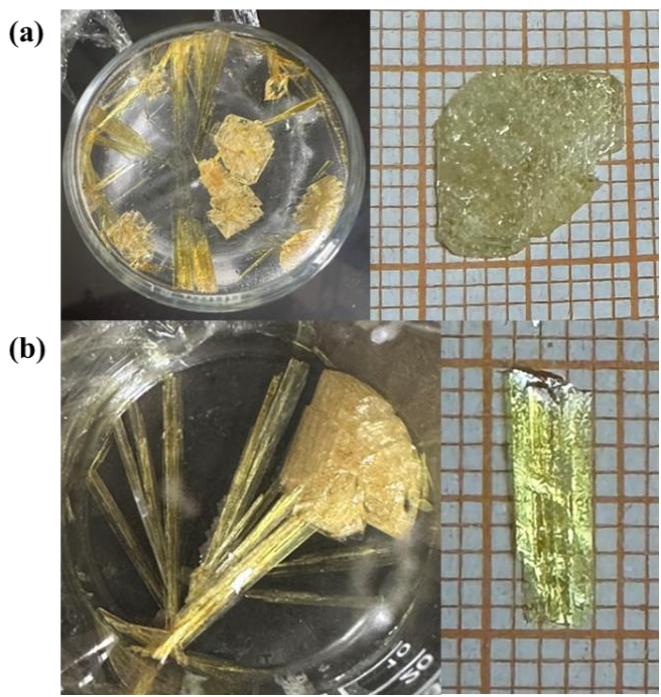

Figure 4 (a) Physical diagram of the (CH$_3$NH$_3$)$_2$HgI$_4$ crystal in the ethyl acetate - toluene system (b) Physical diagram of the (CH$_3$NH$_3$)$_2$HgI$_4$ crystal in the ethyl acetate - chlorobenzene system

The small - sized (CH$_3$NH$_3$)$_2$HgI$_4$ crystals prepared in the experiment were ground into powder samples, and XRD tests were carried out on them. The obtained results were compared with the standard XRD pattern of (CH$_3$NH$_3$)$_2$HgI$_4$ given by the crystal structure database. After comparison, it was found that the XRD pattern of the (CH$_3$NH$_3$)$_2$HgI$_4$ powder prepared in the experiment was highly consistent with the standard pattern, as shown in Figure 5(a). This result clearly proves that the target compound (CH$_3$NH$_3$)$_2$HgI$_4$ was successfully obtained in the experiment, and the preparation process used is effective and feasible. In addition, an X - ray diffraction test was carried out on the flake - shaped crystal in the ethyl acetate - toluene system, and the result is shown in Figure 5(b). The optimal peak in the figure is (204), which belongs to the (102) crystal plane family, consistent with the results of the morphology simulation above. This also indicates that the measured crystal structure information is accurate and can effectively reflect the actual structural characteristics of the material.

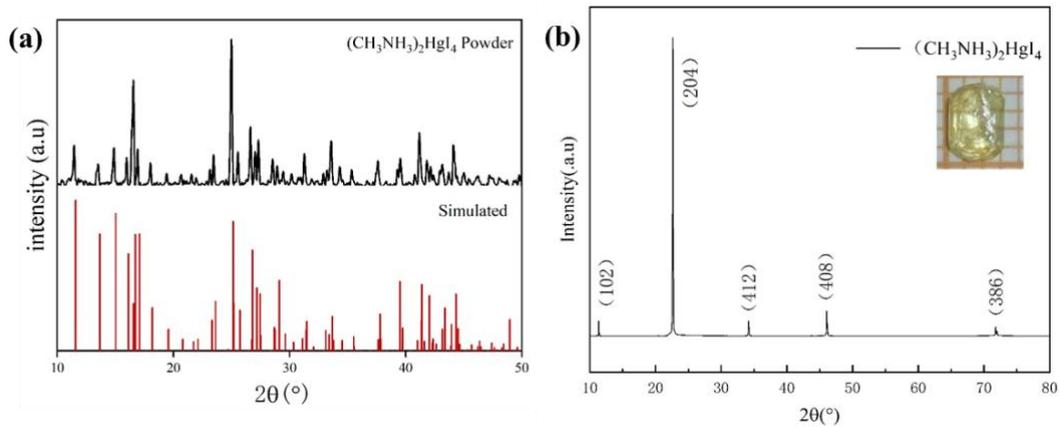

Figure 5 (a) XRD pattern of (CH$_3$NH$_3$)$_2$HgI$_4$ powder (b) XRD pattern of (CH$_3$NH$_3$)$_2$HgI$_4$ crystal

## 4 Discussion

The optical properties of materials are a direct manifestation of the transition behavior between the valence band and the conduction band inside the material after the interaction between light and the material. The study of these properties not only provides a window to deeply understand the physical properties such as the electronic structure and impurity states of the material but also provides a powerful technical means for related research. Its fluorescence spectrum, ultraviolet transmission spectrum, and ultraviolet absorption spectrum were measured through experiments. It is worth noting that the absorption spectra of many low - dimensional hybrid materials at room temperature show obvious exciton absorption characteristics. Therefore, the exciton energy levels and luminescence characteristics of (CH$_3$NH$_3$)$_2$HgI$_4$ were focused on.

The ultraviolet transmission spectrum of (CH$_3$NH$_3$)$_2$HgI$_4$ was tested, as shown in Figure 6. It was observed that its maximum transmittance exceeded 80%. By carefully analyzing the tangent intersection of the spectral data in the linear range with the wavelength axis, the cut-off wavelength of (CH₃NH₃)₂HgI₄ was determined to be approximately 398 nm, and its band-gap value was obtained to be approximately 3.11 eV. This result is close to the band–gap value of 3.28 eV calculated previously using the GGA + U method. A smaller band gap means that the material is more sensitive to photons with longer wavelengths (i.e., lower energy), while a larger band gap corresponds to a response to higher-energy photons. The band-gap value of (CH₃NH₃)₂HgI₄ is approximately 3.11 eV, indicating that this material has a strong photoelectric response ability in the ultraviolet region.

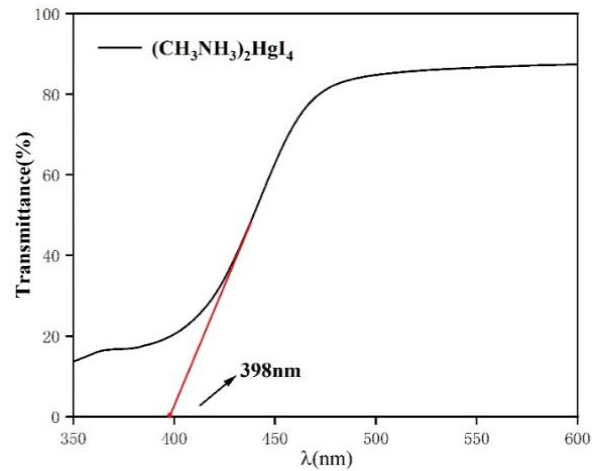

Figure 6 Ultraviolet transmission spectrum of CH₃NH₃)₂HgI₄

The ultraviolet absorption spectrum of (CH₃NH₃)₂HgI₄ was tested, as shown in Figure 7(a). The cut-off wavelength of the (CH₃NH₃)₂HgI₄ crystal is 481 nm. It can be seen that the intrinsic absorption long-wave limit of the (CH₃NH₃)₂HgI₄ crystal is in the visible-blue-green light range. The energy spectrum of the absorption edge was transformed through the following Kubelka-Munk function [9]:

$$(\alpha h\nu)^n = k(h\nu - E_g) \quad (3)$$

Where $\alpha$ is the absorption coefficient, $h\nu$ is the incident photon energy, $k$ is a constant independent of energy, $E_g$ is the band gap, and $n$ is the transition property. For a direct band gap, n=2, and for an indirect band gap, n=1/2. According to the $(\alpha h\nu)^{1/2} \sim h\nu$ plot of (CH₃NH₃)₂HgI₄, the band gap obtained from its linear fit is 2.711 eV, which is smaller than the calculated value of 3.28 eV.

Intrinsic absorption refers to the phenomenon that a material absorbs light through its inherent electronic structure (such as the energy gap between the valence band and the conduction band) in the absence of an external electric or magnetic field. This absorption mainly occurs within the intrinsic band gap of the material, that is, when an electron transitions from one energy level to another, the energy of the photon released or absorbed is equal to the energy difference between these two energy levels. However, the photon energy corresponding to the ultraviolet absorption edge of (CH₃NH₃)₂HgI₄ is less than its band-gap energy, indicating that there may be exciton absorption in this material, and there may be exciton energy levels between the bottom of the conduction band and the top of the valence band.

At room temperature, the absorption spectra of many low-dimensional organic-inorganic hybrid metal halide materials exhibit the characteristics of exciton absorption. The optical absorption peaks of these materials are affected by the strong exciton effect, so it is not appropriate to directly estimate the band-gap width using the optical absorption peaks. Figure 7(b) shows the ultraviolet absorption spectrum of $HgI_2$ and its $(\alpha h\nu)^{1/2} \sim h\nu$ plot. Compared with $HgI_2$, $(CH_3NH_3)_2HgI_4$ has a wider band gap, which plays an important role in photoelectric detection. A larger band gap means that the number of carriers generated by thermal excitation at room temperature is smaller. Therefore, when $(CH_3NH_3)_2HgI_4$ is used as a detector material and operates at room temperature, its resistivity is higher and the leakage current is lower, thus improving the voltage resistance and the maximum operating temperature of the device. $(CH_3NH_3)_2HgI_4$ has a wider band-structure and larger effective masses of electrons and holes, making it highly localized. Low-dimensional organic-inorganic hybrid halides generally exhibit wide band-gaps and highly localized electronic properties. These characteristics endow these materials with extremely high exciton binding energies and stable exciton emissions, which have important application values in fields such as optoelectronic devices, solar cells, and light-emitting diodes.

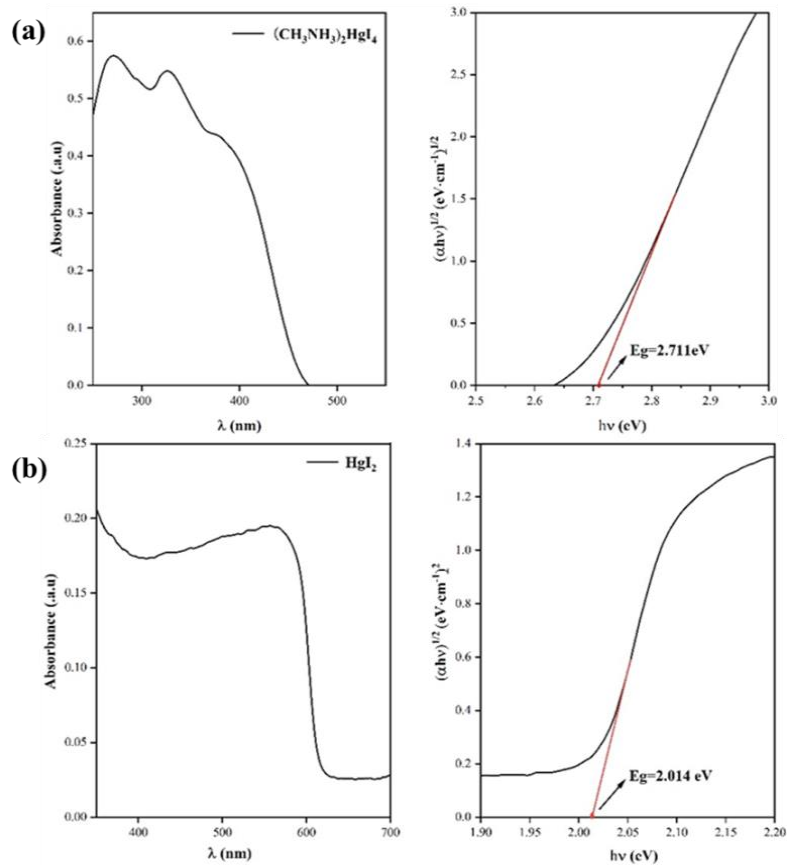

Figure 7(a) Ultraviolet absorption spectrum of $(CH_3NH_3)_2HgI_4$ and its $(\alpha h\nu)^2 \sim h\nu$ plot (b) Ultraviolet absorption spectrum of $HgI_2$ and its $(\alpha h\nu)^2 \sim h\nu$ plot

The existence of excitons in semiconductor materials has a similar physical essence to the electron-proton interaction in hydrogen atoms. Therefore, based on this similarity, the energy

- level formula of hydrogen atoms can be used to simulate and describe the binding energy of excitons [10, 11], as shown in formula (4), which expresses the relationship between the exciton binding energy and material properties (such as dielectric constant, electron charge, and reduced mass, etc.).

$$E_{ex}^n = -\frac{q^4 m_r^*}{32\pi^2 \varepsilon_0^2 \varepsilon_r^2 \hbar^2 n^2} \quad (4)$$

An exciton is a bound - state system formed by the combination of an electron and a hole under the Coulomb interaction. Its energy levels are discrete, similar to the energy levels of hydrogen atoms. When n=1, it represents the ground - state energy level of the exciton, which has the highest binding energy. When n approaches infinity, the exciton binding energy approaches zero, indicating that the electron and the hole are no longer bound to each other. The electron enters the conduction band, and the hole remains in the valence band. For the organic - inorganic hybrid material $(CH_3NH_3)_2HgI_4$, through formula (5)

$$\Delta E_c = \varepsilon_c \frac{\Delta V}{V_0} \quad (5)$$

The calculated exciton binding energy is 302.99 meV when n=1, 75.75 meV when n=2, and 28.39 meV when n=3.

Table 6 Exciton binding energy of $(CH_3NH_3)_2HgI_4$

| n | $E_{ex}^n/meV$ |
| --- | --- |
| 1 | 303.00 |
| 2 | 75.80 |
| 3 | 28.39 |

It was known above that the energy corresponding to the absorption edge of $(CH_3NH_3)_2HgI_4$ is 2.711 eV, and the exciton binding energy is 303.00 meV when n = 1. The total energy obtained by adding them is 3.014 eV, which is close to the experimentally obtained band - gap value of 3.11 eV. This coincidence indicates that the optical transition process in $(CH_3NH_3)_2HgI_4$ is mainly the transition of electrons from the valence band to the n = 1 exciton energy level, and further confirms the stability of the n = 1 exciton energy level in $(CH_3NH_3)_2HgI_4$. This stability is a direct manifestation of the significant exciton characteristics of $(CH_3NH_3)_2HgI_4$ and is also one of the important reasons for its potential application value in the optoelectronic field as a semiconductor material.

Generally, the exciton binding energy is between several tens of millielectron volts and several hundreds of millielectron volts. The exciton binding energy of $(CH_3NH_3)_2HgI_4$ reaches 0.3 eV, indicating that the bound state formed by electrons and holes under the action of electrostatic Coulomb force in the material has high stability. When the exciton binding energy is large, the exciton is more difficult to dissociate and can exist in the material for a longer time. Therefore, it can more effectively participate in the interactions during the spectral reflection and transmission processes. This makes the material show higher sensitivity and detection ability in optical and optoelectronic applications. In addition, a higher exciton binding energy can also improve the luminescence efficiency by reducing the non - radiative recombination of excitons. In InGaN/GaN quantum - well materials, excitons provide energy balance, making the total energy of the exciton system slightly less than the energy of unbound electrons and holes. This energy balance helps to improve the luminescence efficiency [12]. This indicates that by regulating the exciton binding energy, the energy distribution of the material can be optimized, thereby improving its luminescence efficiency.

The luminescence characteristics of $(CH_3NH_3)_2HgI_4$ were further studied through fluorescence spectroscopy experiments. The excitation spectrum of $(CH_3NH_3)_2HgI_4$ was measured at an excitation wavelength of 325 nm, as shown in Figure 8(a). When the excitation wavelength is 325 nm, three significant emission peaks are observed, located at 394 nm, 403 nm, and 438 nm, corresponding to energies of 3.14 eV, 3.07 eV, and 2.83 eV, respectively. The emission peak at 394 nm (3.14 eV) is very close to the experimental band - gap value of 3.11 eV of $(CH_3NH_3)_2HgI_4$, indicating that this peak is its intrinsic emission peak. In contrast, the energy differences between the emission peaks at 403 nm (3.07 eV) and 438 nm (2.83 eV) and the intrinsic emission peak are 0.07 eV and 0.31 eV, respectively. These differences are in good agreement with the exciton binding energies of $(CH_3NH_3)_2HgI_4$ when n = 1 and n = 2 calculated previously (302.99 meV and 75.75 meV, respectively). This result reveals that these two emission peaks are exciton emission peaks, further confirming that $(CH_3NH_3)_2HgI_4$ has stable excitons at room temperature.

The fluorescence spectra of $(CH_3NH_3)_2HgI_4$ were measured at excitation wavelengths of 325 nm, 330 nm, and 340 nm. As can be seen in Figure 8(b), the positions of the emission peaks shift at different excitation wavelengths, showing excitation - wavelength dependence. When a substance is photo - excited, its electrons transition from the ground state to the excited state. During the process of returning from the excited state to the ground state, the electrons interact with the surrounding atoms or molecules. This energy transfer causes the wavelength of the emitted light to be longer than that of the excitation light, that is, a red - shift occurs. This property has a wide range of applications in many fields, including imaging, optical anti - counterfeiting, biological labeling, and optoelectronic devices. For example, carbon quantum dots (CDs) exhibit unique excitation - dependent fluorescence characteristics. That is, as the excitation wavelength changes, their emission wavelength also changes accordingly. By adjusting the excitation wavelength, continuous full - color emission of CDs can be achieved, which provides possibilities for their application in the multi - color imaging biological field [13]. In addition, hybrid rare - earth - based halides also exhibit efficient excitation - wavelength - dependent fluorescence characteristics. The positions of their emission peaks change with the change of the excitation wavelength, making them have broad application prospects in anti - counterfeiting, biological labeling, and optoelectronic devices [14].

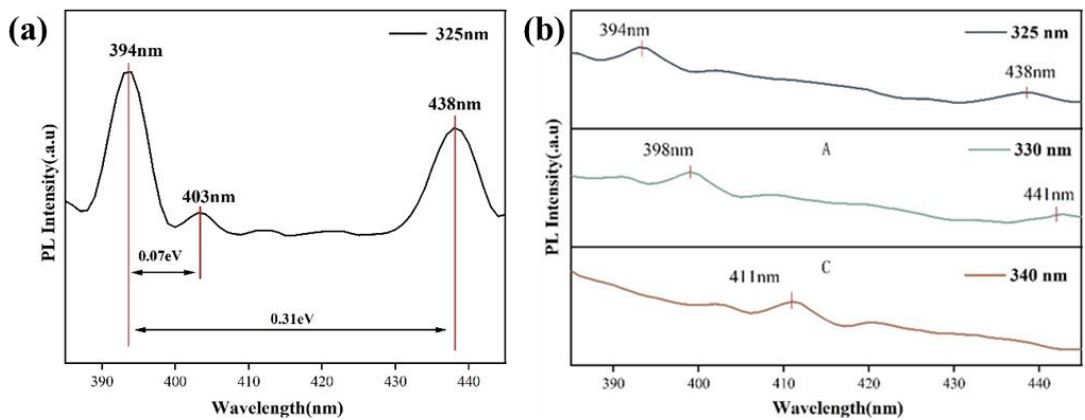

Figure 8 (a) Exciton luminescence spectrum of $(CH_3NH_3)_2HgI_4$ (b) Fluorescence spectrum of $(CH_3NH_3)_2HgI_4$

## 5 Conclusion

This paper studied the influence of solvents and anti - solvents on the crystal morphology, and thus selected a solvent - anti - solvent system suitable for large - area/volume crystal growth. Then, relatively large – sized $(CH_3NH_3)_2HgI_4$ crystals were successfully grown by the anti - solvent method. The band structure and electronic structure of $(CH_3NH_3)_2HgI_4$ were simulated and calculated by first - principles, and its optical properties were comprehensively analyzed by combining experimental tests. The main conclusions are as follows: The morphology simulation results of the $(CH_3NH_3)_2HgI_4$ crystal show that the ethyl acetate - toluene system is more suitable for the growth of large - area/volume $(CH_3NH_3)_2HgI_4$ crystals. A flake - shaped crystal with a volume of approximately 10 mm×9 mm×1 mm was successfully grown by the anti - solvent method. XRD diffraction analysis verified that the preparation process used in the experiment is effective and feasible, and the largest exposed plane belongs to the (102) crystal - plane family, which is consistent with the morphology simulation results. The exciton binding energy levels of $(CH_3NH_3)_2HgI_4$ were calculated. The exciton binding energy of the first energy level is 303.00 meV. It has a large exciton binding energy, stable exciton energy levels, and significant exciton characteristics. The exciton luminescence and intrinsic luminescence of $(CH_3NH_3)_2HgI_4$ were studied by a fluorescence spectrophotometer. Under the excitation of a 325 - nm - wavelength excitation light at room temperature, the results show that $(CH_3NH_3)_2HgI_4$ has an intrinsic emission peak at 394 nm, and the emission peaks at 403 nm and 438 nm are exciton peaks. The corresponding energy differences are in good agreement with the calculated exciton binding energies, confirming the exciton characteristics of $(CH_3NH_3)_2HgI_4$, the existence of exciton luminescence, and its excitation - wavelength dependence.